\documentclass[12pt,preprint]{aastex}

\slugcomment{To appear in The Astrophysical Journal Letters}
\shorttitle{H$_2$CO Maser in G$\,$23.71$-$0.20}
\shortauthors{Araya, et al.}

\def \kms {$\,$km s$^{-1}$}
\def \lsols {$\,$L$_{\odot}$}

\def \mjyb {$\,$mJy$\,$beam$^{-1}$}
\def\fsecs{\hbox{$.\!\!^{\rm s}$}}

\begin{document}

\title{A New Galactic 6$\,$cm Formaldehyde Maser}

\author{E. Araya \altaffilmark{1},
P. Hofner\altaffilmark{1,}\altaffilmark{2},
W. M. Goss \altaffilmark{2},
S. Kurtz\altaffilmark{3},
H. Linz\altaffilmark{4},
L. Olmi\altaffilmark{5,}\altaffilmark{6}}

\altaffiltext{1}{New Mexico Institute of Mining and Technology, 
Physics Department, 801 Leroy Place,
Socorro, NM 87801.}
\altaffiltext{2}{National Radio Astronomy Observatory, P.O. 
Box 0, Socorro, NM 87801.}
\altaffiltext{3}{Centro de Radioastronom\'{\i}a y Astrof\'{\i}sica, UNAM,
Apdo. Postal 3-72, 58089, Morelia, Michoac\'an, M\'exico.}
\altaffiltext{4}{Max--Planck--Institut f\"ur Astronomie, K\"onigstuhl 17,
D--69117 Heidelberg, Germany.}
\altaffiltext{5}{Istituto di Radioastronomia, CNR, Sezione di Firenze,
Largo Enrico Fermi 5, I-50125 Florence, Italy.}
\altaffiltext{6}{University of Puerto Rico at Rio Piedras, Physics
Department, P.O. Box 23343, San Juan, PR 00931.}

\begin{abstract}
We report the detection of a
new H$_2$CO maser in the massive star forming region 
G23.71$-$0.20 (IRAS$\,$18324$-$0820), i.e., the fifth
region in the Galaxy where H$_2$CO maser emission has been
found.
The new H$_2$CO maser is located toward a compact H{$\,$\small II}
region, and is coincident in velocity and position with 6.7$\,$GHz methanol
masers and with an IR 
source as revealed by Spitzer/IRAC GLIMPSE data.
The coincidence with an IR source and 6.7$\,$GHz methanol
masers suggests that the maser is in close proximity to
an embedded massive protostar.
Thus, the detection of H$_2$CO maser emission toward 
G23.71$-$0.20 supports the trend that H$_2$CO 6$\,$cm masers trace 
molecular material very near young massive stellar objects.
\end{abstract}


~
\newpage

\section{Introduction}

The 6$\,$cm transition of formaldehyde (H$_2$CO, 
J$_{K_a K_c} = 1_{11} - 1_{10}$) was among the first molecular
lines detected in the interstellar medium.
This transition is ubiquitously observed in absorption against Galactic
continuum sources (e.g., Araya et al. 2002) as 
well as against the 2.7$\,$K Cosmic Microwave Background 
(e.g., Palmer et al. 1969). 
In contrast, only a handful of H$_2$CO 6$\,$cm emitters have been 
detected: H$_2$CO megamaser emission has 
been confirmed toward three galaxies (Araya, Baan \& Hofner 2004a), and 
in our Galaxy H$_2$CO emission has been detected toward 5 sources:
(quasi) thermal emission toward the Orion BN/KL region, and 
maser emission toward NGC$\,$7538 IRS1, Sgr B2, G29.96$-$0.02,
and recently IRAS$\,$18566+0408 (Araya et al. 2005).

All Galactic emitters are found in
close proximity to signposts of young massive stars,
indicating that the physical conditions
necessary for H$_2$CO 6$\,$cm emission occur in
early phases of massive star formation. In addition, the low
detection rate of H$_2$CO masers may be a consequence of 
specific and short lived physical conditions in massive star 
forming regions, thus H$_2$CO masers could become a valuable 
astrophysical probe if the excitation
mechanism of the maser was known. However,
the physical mechanism for H$_2$CO 6$\,$cm maser emission is 
still not understood. Currently, the only quantitative model 
proposed to explain H$_2$CO 6$\,$cm masers is that of 
Boland \& de Jong (1981),
where the level inversion is caused by the 
radio continuum from a background compact H{\small II}
region.
Unfortunately, this model cannot explain most of the known
H$_2$CO 6$\,$cm masers (e.g., Hoffman et al. 2003).
The lack of progress in the understanding of the H$_2$CO
6$\,$cm maser mechanism is in part due to the small
sample of known H$_2$CO maser regions. A larger sample of 
H$_2$CO maser regions would
not only serve to further test the Boland \& de Jong (1981)
model, but also to investigate the dependence of
H$_2$CO masers on a larger variety of physical environments, 
e.g., to check whether H$_2$CO masers are preferentially associated with 
shocked gas or with more quiescent molecular material that could be
radiatively excited.

With the goal of increasing the number of known 
H$_2$CO 6$\,$cm maser regions, 
we conducted in 2002 and 2003 a survey for H$_2$CO 
6$\,$cm emission with the Arecibo telescope (Araya et al. 2004b).
In the survey we observed massive star forming regions
characterized by high molecular densities and 
low radio-continuum at 6$\,$cm, and detected  
H$_2$CO 6$\,$cm maser emission toward IRAS$\,$18566+0408
(Araya et al. 2005). Motivated by our first Arecibo survey, 
we recently completed a second survey with the GBT and 
VLA\footnote{The 100$\,$m Green Bank Telescope (GBT) 
and the Very Large Array (VLA) are operated by 
the National Radio Astronomy Observatory (NRAO), 
a facility of the National Science Foundation operated 
under cooperative agreement by Associated Universities, Inc.}. 
The main result of our second survey,
which is reported in this letter, is the discovery 
of a new H$_2$CO 6$\,$cm maser toward the massive star forming region
G23.71$-$0.20.

\section{Observations and Data Reduction}

On 2005 January 10, we conducted an H$_2$CO 6$\,$cm maser survey toward 
10 massive star forming regions with the VLA in the A configuration.
The sources were selected based on single dish 
H$_2$CO spectra that had been obtained by our group with the 
Arecibo and GBT telescopes (Watson et al. 2003, Sewilo et al. 2004a), 
and that were consistent with H$_2$CO absorption blended 
with emission. 
We detected radio continuum and H$_2$CO 6$\,$cm 
absorption toward several regions, and a new H$_2$CO 6$\,$cm emitter toward
G23.71$-$0.20 (IRAS$\,$18324$-$0820). This latter detection is
the topic of this letter. The results for the non-emitting regions 
will be presented in a future paper (Araya et al. {\it in prep.}). 
Further observations of the G23.71$-$0.20 region were conducted
on 2005 April 24, with the VLA in the B configuration. 
These observations
were intended to confirm the detection.
In Table~1, we list details of the spectral line observations conducted 
with the VLA in the A and B configurations. 
The data were calibrated and imaged in AIPS following standard 
spectral-line reduction procedures. Only external calibration (i.e., no 
self-calibration) was used to obtain the complex gain corrections for
data calibration. We did not detect radio continuum in individual
channels, hence continuum subtraction was not necessary. 
No bandpass calibration was necessary due to the narrow bandwidth.

\vspace{-0.7cm}
\section{Results and Discussion}

\subsection{A New H$_2$CO 6$\,$cm Maser in the Galaxy}

Using the VLA in the A configuration we detected H$_2$CO 6$\,$cm 
emission toward G23.71$-$0.20.
The emission was detected in two channels (i.e., width 0.76\kms) with 
an rms noise of 4.5\mjyb$\,$ in a natural weighted map. The peak intensity is 
I$_{\nu} =\,$60\mjyb~ at the velocity V$_{max} =\,$79.2\kms. 
The maximum intensity of the H$_2$CO emission
is located at $\alpha$(J2000) = 18$^{\rm h}$35$^{\rm m}$12\fsecs37, 
$\delta$(J2000) = $-$08\arcdeg 17\arcmin39\farcs3. 
The emission feature is unresolved ($\theta_{source} <~$0\farcs4), 
which implies a lower limit on the brightness 
temperature of $\sim$30000$\,$K, i.e., the emission is due to a 
maser mechanism. 
The intensity of this new maser is similar to the
70$\,$mJy/beam of the maser in G29.96$-$0.02 (Pratap, Menten, 
\& Snyder, 1994).
On the other hand, the maser in G23.71$-$0.20 is narrower in comparison
with other H$_2$CO masers (e.g., 
the FWHM of the maser in IRAS$\,$18566+0408 
is 1.6\kms, Araya et al. 2005, whereas the FWHM of the new
maser is smaller than 0.8\kms).

The VLA-B observations confirmed the existence of the 
H$_2$CO 6$\,$cm maser in G23.71$-$0.20. 
As in the VLA-A observations, the maser is detected
in two channels (0.76\kms) with a signal to noise ratio
greater than 4 (rms $\sim$ 6.0\mjyb).
We measured a peak velocity and intensity values of V$_{max} = 79.2$\kms~
and I$_{\nu} =\,$44\mjyb~(T$_B > 400\,$K), respectively. 
The H$_2$CO peak intensity measured with the VLA-B is less than 
the peak intensity measured with the VLA-A, however both values
are consistent within 5$\sigma$.
We combined the VLA-A and B {\it uv} data to produce the
map and spectrum shown in Figure~1. The spectrum in Figure~1 shows that
H$_2$CO emission may also be present in the two lower-velocity
channels blueward of the peak channel.

\vspace{-0.7cm}
\subsection{An Overview of the G23.71$-$0.20 Massive Star Forming Region}

G23.71$-$0.20 is a massive star forming region located in the 
Scutum constellation. Sewilo et al. (2004a) reported H110$\alpha$ 
emission from the
region at a LSR velocity of 76.5\kms, hence,
the H$_2$CO maser LSR velocity is coincident 
(within $\sim$2.7\kms)
with the H110$\alpha$ line center. This velocity correspondence 
implies that the H$_2$CO maser is associated with the
G23.71$-$0.20 massive star forming region.
Based on the LSR velocity of the H110$\alpha$ line, 
the two possible kinematic distances of G23.71$-$0.20 
are 4.9$\,$kpc and 11$\,$kpc. 
Sewilo et al. (2004a) considered the far kinematic distance 
(i.e., D$_{LSR} \sim 11\,$kpc) as more likely.

Becker et al. (1994) (see also 
White, Becker, \& Helfand 2005) 
detected 6$\,$cm radio continuum from the region. 
In Figure~2 we show their VLA-C 6$\,$cm 
continuum map\footnote{Multi-Array Galactic Plane 
Imaging Survey, http://third.ucllnl.org/gps/.}. 
As the figure illustrates, the H$_2$CO maser is not coincident
with the strong radio continuum emission, but rather lies in an area of
more diffuse emission. Apart from the 6$\,$cm detection by 
White et al. (2005), no other high-angular resolution 
detection of the radio continuum at frequencies above 
2$\,$GHz\footnote{The region was
detected in VLA L-band (20$\,$cm) surveys: NVSS (Condon, et al. 1998), and 
the Multi-Array Galactic Plane Imaging Survey (White et al. 2005).}
is available. Future high-angular resolution
observations of the radio continuum of the region at several
frequencies are required to test if the H$_2$CO maser
in G23.71$-$0.20 can be explained by the Boland \& de Jong (1981)
model.

The radio continuum source in G23.71$-$0.20 is coincident with 
IRAS$\,$18324$-$0820, which has the characteristic 
far-infrared color of ultra-compact H {\small II} regions 
(Sewilo et al. 2004a). 
Assuming a distance of 11$\,$kpc, isotropic emission, 
and following the formulation of 
Casoli et al. (1986), we estimate a bolometric luminosity of 
$\sim 2.4\times 10^5$\lsols, which corresponds to the luminosity of 
an $\sim$O6 ZAMS star (Panagia 1973). 

High angular resolution studies of molecular line transitions
toward G23.71$-$0.20
are as scarce as the radio continuum observations. 
The only available interferometric molecular data 
of the region are from Walsh et al. (1998). They 
detected seven CH$_3$OH 6.7$\,$GHz maser spots in the region. 
In Figure~2, we show the positions of these masers. 
The CH$_3$OH 6.7$\,$GHz masers are all coincident with the H$_2$CO
6$\,$cm maser within $\sim$1$\arcsec$, which is the absolute positional 
accuracy of the CH$_3$OH masers (Walsh et al. 1998). 
The masers are spread over a 
velocity range from 74.9 to 81.4\kms, which encompasses
the velocity of the H$_2$CO maser (79.2\kms).

G23.71$-$0.20 has been detected in 
several single dish surveys for 
CH$_3$OH masers:
Blaszkiewicz \& Kus (2004) (12.2$\,$GHz CH$_3$OH masers); 
Schutte et al. (1993), Slysh et al. (1999), Szymczak, Hrynek, \& Kus (2000)
(6.7$\,$GHz CH$_3$OH masers). In all cases,
the CH$_3$OH masers show similar velocities to that of the 
H$_2$CO maser. G23.71$-$0.20 was also observed in CS J=2-1
by Bronfman, Nyman, \& May (1996). They detected CS J=2-1 emission
at a velocity of 68.3\kms, which is 
approximately 10\kms~lower than that of the H$_2$CO maser. 
Finally, Han et al. (1998) conducted a survey of
22$\,$GHz H$_2$O masers at the Purple Mountain
Observatory, and report detection of a 35.6$\,$Jy H$_2$O maser 
at a LSR velocity of $-$40.3\kms. This radial velocity is 
very different from the velocity of the H$_2$CO
maser (79.2\kms), thus the association of this H$_2$O
maser with the H$_2$CO and CH$_3$OH masers is unclear.

We searched the MSX and Spitzer/IRAC GLIMPSE images for infrared (IR)
emission in the vicinity of the H$_2$CO maser position. The
21.3 $\mu$m emission seen in the MSX E band (angular resolution
18$''$) partly traces the two main 6$\,$cm continuum emission regions
to the SE and SW of the maser position, but also shows extended 
emission near the maser position.  
The Spitzer/IRAC GLIMPSE data show
a compact IR source located within $0\farcs3$ of the H$_2$CO maser position.
This source is detected in all four IRAC bands.
In Figure~3 we show a 
color--coded image of the Spitzer data (blue is the average of the 
3.6 and 4.5 $\mu$m data, green is 5.8 $\mu$m, and
red is 8.0 $\mu$m), and we mark the position of the H$_2$CO maser with a cross.
While the [3.6]--[4.5] color of 2.1 is indicative of a deeply embedded object,
the [5.8]--[8.0] color of 0.1 is peculiar in combination with the first
color (see Indebetouw et al. 2006). 
The [5.8]--[8.0] color could be caused by unresolved source 
multiplicity and/or the presence of PAH emission features. 
In addition, strong and broad $\sim$10 $\mu$m silicate absorption
could be affecting the 8.0 $\mu$m band.
High angular resolution infrared observations are underway 
to further study this IR source.

\subsection{Implication for the Nature of H$_2$CO 6$\,$cm Masers}

The inversion mechanism of H$_2$CO 6$\,$cm masers is 
unclear (e.g., Mehringer, Goss, \& Palmer 1994). Besides 
the model by Boland \& de Jong (1981) which is based on radiative 
excitation by radio continuum, Hoffman et al. (2003) suggested 
that H$_2$CO 6$\,$cm masers could be collisionally pumped.
The new detection of an H$_2$CO
6$\,$cm maser in G23.71$-$0.20 and the coincidence of the 
maser with a Spitzer compact source (Figure~3) suggests 
another possibility: H$_2$CO 6$\,$cm masers could be radiatively 
pumped by IR photons in massive star forming regions.
A connection between FIR radiation and H$_2$CO 6$\,$cm
masers has been noticed before in megamaser galaxies, e.g.,
the data by Araya et al. (2004a) suggest a 
correlation between the luminosity of H$_2$CO megamaser
lines and FIR luminosity (see also Baan, Haschick, \& 
Uglesich 1993). In addition, Litvak (1970) pointed out 
that absorption of infrared radiation in conjunction with
large H$_2$CO optical depths and H$_2$CO density comparable to
that of OH maser regions, may result in H$_2$CO maser emission.
Quantitative models have been
developed to explain OH masers via FIR radiative excitation
(e.g., Henkel, G\"usten, \& Baan 1987; Cesaroni \& Walmsley
1991), however, to date there exist no
equivalent models exploring pumping of the observed H$_2$CO
6$\,$cm masers via IR photons.

The idea of FIR pumping of H$_2$CO masers is compelling;
however a quantitative test of this hypothesis is 
not practical since (as mentioned in section 3.2) the 
available data toward the H$_2$CO maser in G23.71$-$0.20 
do not provide sufficient constraints to carry 
out an analysis of the inversion mechanism. Specifically,
the FIR fluxes measured by IRAS toward G23.71$-$0.20 are
likely to trace the extended IR region to the SW
of the maser position (see Figure~3) and not the FIR
properties of the compact Spitzer source.

In the case of G23.71$-$0.20, the other proposed 
pumping mechanisms cannot be currently tested either.
The radio continuum at the position of the maser is blended 
with emission from two nearby continuum regions, thus
neither the gain of the maser or the emission
measure of the ionized gas at the maser position can
be determined.
In addition, the thermal molecular data (e.g., CS) are insufficient to 
establish whether the H$_2$CO maser is associated with a molecular clump 
or outflow, and we also lack high quality information 
about shock tracers like H$_2$O masers that may be 
associated with H$_2$CO masers (e.g., Araya et al. 2005). 

On the other hand, the detection of maser emission
in G23.71$-$0.20 supports the trend that
H$_2$CO masers reside very near to massive young stellar
objects (YSOs), i.e., closer than a few thousand AU.
For example, the maser in G29.96$-$0.02 is coincident with a 
hot molecular core (Pratap et al. 1994) that probably 
contains a massive circumstellar disk (Olmi et al. 2003); 
the maser in IRAS$\,$18566+0408 
is coincident with a 2MASS and a bright Spitzer/GLIMPSE source 
(Araya et al. 2005, 2006 in prep.) and the central continuum
object has been classified as a massive disk candidate (Zhang 2005).
Also, the maser in NGC$\,$7538 IRS1 (which shows an H$_2$CO NE$-$SW 
velocity gradient, Hoffman et al. 2003) is located at the
position of an hypercompact H{$\,$\small II} region candidate
(Sewilo et al. 2004b) that harbors a possible circumstellar 
disk oriented in a NE$-$SW direction (De Buizer \& Minier 2005). 
The close association of H$_2$CO masers with massive
YSOs is in contrast to other molecular masers 
(e.g., 44$\,$GHz CH$_3$OH masers, Kurtz, Hofner, 
\& Vargas-\'Alvarez 2004) 
that can be found at larger 
distances from massive YSOs and are probably
related to the interaction of jets and outflows with 
the surrounding material of the molecular cloud.
If the H$_2$CO masers are associated with shocked gas,
then they might trace the regions where accretion from 
a mass reservoir to a massive circumstellar disk occurs.

\section{Summary}

Using the VLA in the A configuration we detected H$_2$CO
emission from the massive star forming region G23.71$-$0.20.
Based on the brightness temperature limit of the detection
(T$_b > 30000\,$K), the 
line must be due to a maser mechanism: i.e., this source is the
fifth region in the Galaxy where H$_2$CO 6$\,$cm maser emission has 
been detected. The FWHM of the line is $< 0.8$\kms, i.e., narrower
than other H$_2$CO masers detected with the VLA.
The maser was independently confirmed by VLA-B observations 
of the region, that were conducted approximately 3 months
after the initial detection. 
The LSR velocity and position of the maser closely correspond 
to those of CH$_3$OH 6.7$\,$GHz masers detected by Walsh et al. (1998) 
with the Australia Telescope Compact Array. 
We found a compact Spitzer/IRAC IR source, possibly a deeply 
embedded young stellar object, coincident with the H$_2$CO maser. 
The detection of H$_2$CO maser emission in G23.71$-$0.20 supports 
the trend that H$_2$CO 6$\,$cm masers are located very near
massive YSOs.

\acknowledgments

Support for this work was provided by the NSF through award 
GSSP 05-0006 from the NRAO. 
Part of the observations presented in this paper
were conducted as part of a VLA student project directed by   
D. Westpfahl at NMT.
P. H. acknowledges support from 
NSF grant AST-0454665. H. L. was supported by a Postdoc stipend of the German
Max Planck Society. We also acknowledge an anonymous referee
for comments that improved the manuscript.
This research has made use of
NASA's Astrophysics Data System and is based in part on observations made
with the Spitzer Space Telescope, operated by JPL, CalTech under contract
with NASA.

\begin{deluxetable}{lcc}
\tabletypesize{\scriptsize}
\tablecaption{VLA Observations \label{tbl-1}}
\tablewidth{0pt}
\tablehead{
\colhead{Parameter} & \colhead{VLA-A} & 
\colhead{VLA-B} }
\startdata
\rm Date           & 2005 Jan. 10                      & 2005 Apr.  24           \\ 
RA$^a$	           & 18$^{\rm h}$35$^{\rm m}$12\fsecs54            & 18$^{\rm h}$35$^{\rm m}$12\fsecs54 \\
Dec$^a$            & $-$08\arcdeg17\arcmin46\farcs5    & $-$08\arcdeg17\arcmin46\farcs5\\
$\nu_o\,$(GHz)$^b$ & 4829.6569                         & 4829.6569 \\
V$_{LSR}$(\kms)$^c$& 80.0                              & 80.0      \\
IF Mode            & 2IF (AD)                          & 2IF (AD)  \\
BW~(MHz)$^d$       & 1.56                              & 1.56      \\
~~~~~~(\kms)       & 97.0                              & 97.0      \\
$\Delta\nu^e$ (kHz)& 6.104                             & 6.104     \\
~~~~~~(\kms)       & 0.38                              & 0.38      \\
Synth. Beam        & 0\farcs63$\times$0\farcs41        & 3\farcs85$\times$1\farcs44\\
Position Angle     & 1\arcdeg                          & 49\arcdeg \\
Flux Density Calib.& 3C$\,$286                         & 3C$\,$48  \\
~~~~~~Assumed S$_\nu$ & 7.52                              & 5.47      \\
Phase Calib.       & J1832$-$105                         & J1832$-$105 \\
~~~~~~Measured S$_\nu$& 1.3                               & 1.2      \\
rms (mJy/beam)$^f$ & 4.5                               & 6.0       \\
\enddata
\tablenotetext{a}{Phase tracking center (J2000) of the G23.71$-$0.20 observations.}
\tablenotetext{b}{Rest frequency of the H$_2$CO J$_{K_aK_c} = 1_{10} - 1_{11}$ transition.}
\tablenotetext{c}{Central bandpass velocity.}
\tablenotetext{d}{Bandwidth per IF.}
\tablenotetext{e}{Final channel width.}
\tablenotetext{f}{Final rms in a channel map.}
\end{deluxetable}

\clearpage
\begin{figure}

~\\
\vspace{13cm}
\includegraphics{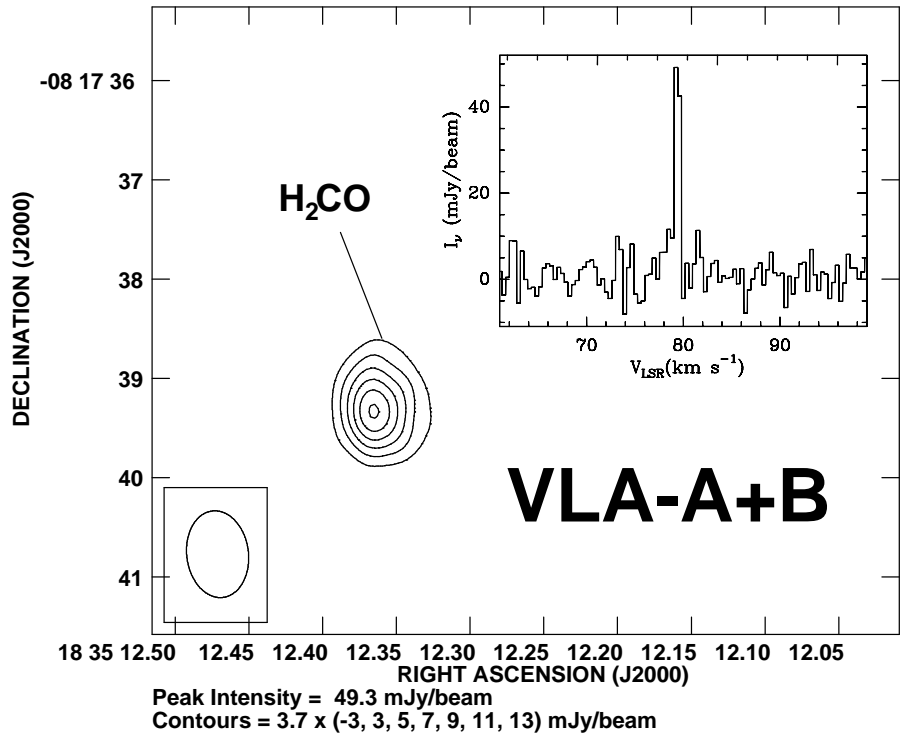}  
\figurenum{1}
\caption{
H$_2$CO 6$\,$cm maser emission toward
G23.71$-$0.20. We show the peak channel image (contours) 
and the H$_2$CO spectrum of the VLA-A+B combined data. 
The rms of the image is 3.7\mjyb, and $\theta_{syn}$ 
is 0\farcs88$ \times 0\farcs62$, P.A. = 7$\arcdeg$.}
\end{figure}

\clearpage
\begin{figure}

~\\
\vspace{11cm}
\includegraphics{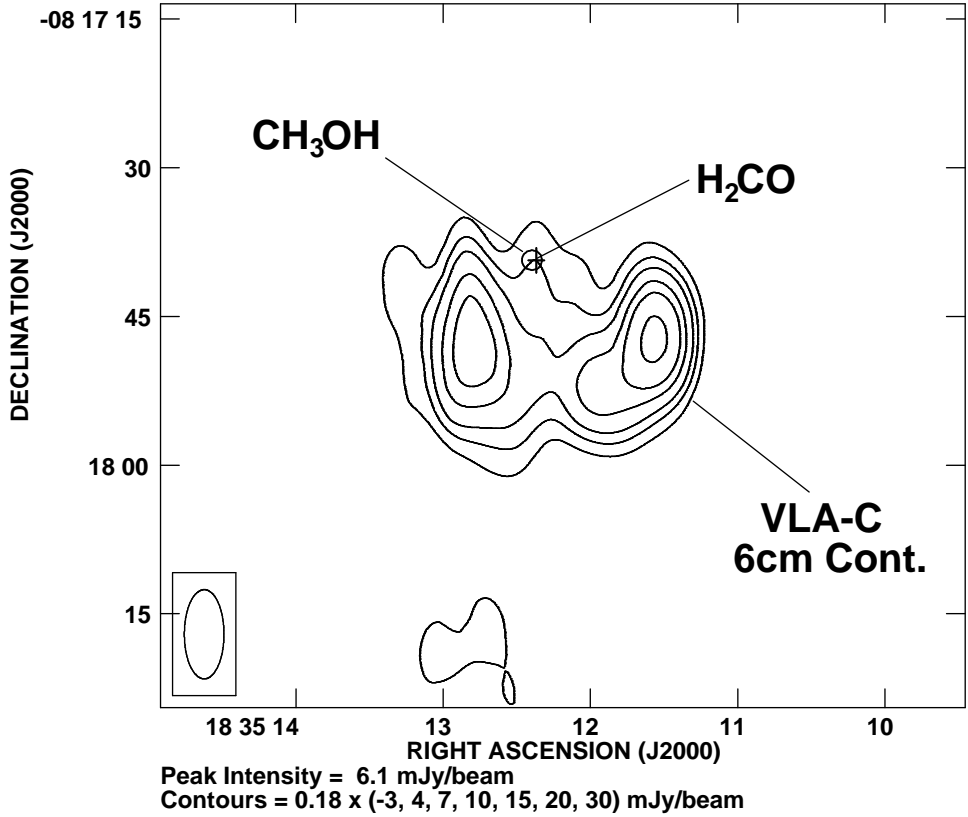}  
\figurenum{2}
\caption{6$\,$cm VLA-C map from 
the Multi-Array Galactic Plane Imaging Survey 
(http://third.ucllnl.org/gps/, White et al. 2005).
The rms of the image is 0.18\mjyb, the peak intensity is
6.1\mjyb, and the synthesized beam is $\sim 9\arcsec \times 4\arcsec$,
P.A. $\sim$ 0\arcdeg. 
We also show the position of the H$_2$CO maser
with a plus symbol (the size of the symbol is four times the size of the
VLA-A synthesized beam),
and the location of seven 6.7$\,$GHz CH$_3$OH masers reported by 
Walsh et al. (1998) with a circle (all of them are located
within 0\farcs2 from each other, and the size of the circle is 
approximately twice the absolute position accuracy of
the CH$_3$OH masers).} 
\end{figure}

\clearpage
\begin{figure}

~\\
\vspace{15cm}
\includegraphics{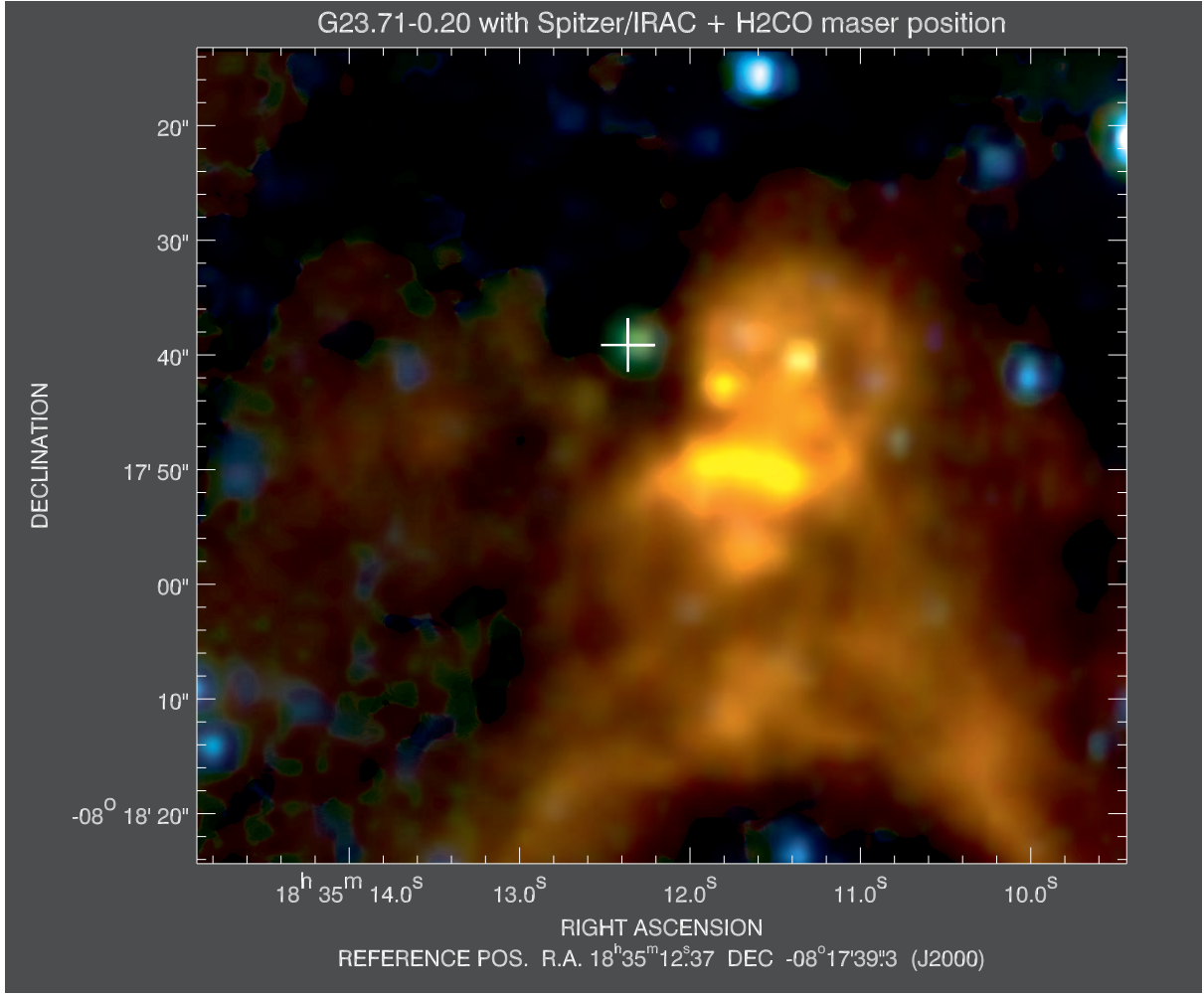}  
\figurenum{3}
\caption{
Spitzer/IRAC GLIMPSE color composite of the G23.71--0.20 region. Blue is the
average of the 3.6 and 4.5 $\mu$m data, green is 5.8 $\mu$m, and red is
8.0 $\mu$m. The cross marks the position of the H$_2$CO maser. Note the
detection of a compact IR source at the maser position.}
\end{figure}

\end{document}